\documentclass[amssymb,pra,twocolumn,notitlepage,superscriptaddress]{revtex4-1}

\renewcommand{\vec}[1]{\mathbf{#1}}

\newcommand{\braket}[2]{\langle #1 | #2 \rangle}
\newcommand{\innerproduct}[2]{\langle #1, #2 \rangle}

\newcommand{\ket}[1]{| #1 \rangle}
\newcommand{\bra}[1]{\langle #1|}

\usepackage{color} 

\usepackage{amsmath}                          % better math control.
\usepackage{amsfonts}                         % extended math fonts.
\usepackage{amssymb}                          % names amsfont symbols.
\usepackage{amsthm}                           % theorem environments
\usepackage{mathdots}                         % for ddots going upwards 

\usepackage{graphicx}                         % ps, pdf figs.
\usepackage{subfigure}                        % for subfigures
\usepackage{overpic}

\begin{document}

%%%%%%%%%%%%%%%%%%%%%%%%%%%%%%%%%%%%%%%%%%%%%%%%%%%%%%%%%%%%%%%%%%%%%%%%%%%%%%%

\title{Nonlinear regression based on a hybrid quantum computer}
% hybrid quantum 
%  Continuous-variable assi
\author{Dan-Bo Zhang}
%\email{dbzhang@m.scnu.edu.cn} 
\affiliation{Guangdong Provincial Key Laboratory of Quantum Engineering
	and Quantum Materials, and School of Physics\\ and Telecommunication Engineering,
	South China Normal University, Guangzhou 510006, China}

\author{Shi-Liang Zhu} \email{slzhu@nju.edu.cn}
\affiliation{Guangdong Provincial Key Laboratory of Quantum Engineering
	and Quantum Materials, and School of Physics\\ and Telecommunication Engineering,
	South China Normal University, Guangzhou 510006, China}
\affiliation{National Laboratory of Solid State Microstructures, School of Physics, Nanjing University, Nanjing 210093, China} 

\author{Z. D. Wang}
\email{zwang@hkucc.hku.hk}
\affiliation{Department of Physics and Center of Theoretical and Computational Physics, The University of Hong Kong, Pokfulam Road, Hong Kong, China}

\begin{abstract}
Incorporating nonlinearity into quantum machine learning is essential for learning a complicated input-output mapping. We here propose quantum algorithms for nonlinear regression, where nonlinearity is introduced with feature maps when loading classical data into quantum states. Our implementation is based on a hybrid quantum computer, exploiting both discrete and continuous variables, for their capacity to encode novel features and efficiency of processing information. We propose encoding schemes that can realize well-known polynomial and Gaussian kernel ridge regressions, with exponentially speed-up regarding to the number of samples.

% Encoding vector into quantum states. It casts kernel as inner product of quantum states. 
%Rather than just considering the input step as a bottleneck for quantum computing, we exploit the opportunity it gives by investigating properties of the encoding of classical information into quantum states. We find that the encoding naturally introduce kernels.
\end{abstract}

% to-to-list
% delect the extension. Leave it for further work. 
% physical realization in iron trap.
% prepare the matrix state without using quantum random access memory.

\maketitle
% normalization of quantum states
% run time analysis, after quantum linear regression has been dicsussed. 
\section{Introduction}
%To write it in a direct way.
%\emph{Motivation}

Machine learning is renowned for its power of pattern recognition~\cite{christopher_m_bishop_pattern_2006,trevor_hastie_elements_2009}. It learns a mapping between a high dimensional input and a much simpler output such as a discrete label for classification or a continuous variable for regression. For many real-world applications, the mapping may be complicated. To establish a fine input-output relation, a simple linear model is probably not well applicable. To tackle this problem, one may refer to a proper feature map that generates new features, and the task may be still well solved by a simple (linear) model in this new feature space. Moreover, kernel methods~\cite{christopher_m_bishop_pattern_2006} can be applied without explicitly making time-consuming feature maps.  %Examples\cite{trevor_hastie_elements_2009} include kernel ridge regression and kernel support vector machines(SVM). The former, for instance, has been used to predict energy solely from the structure of molecules without solving density functional calculation explicitly using Gaussian kernel.

% some theoretical papers on enhanced QML should be cited
Recent efforts on machine learning have been devoted to exploiting the capability of quantum computing~\cite{biamonte_17,wiebe_12,schuld_15,lloyd_14,rebentrost_14,dunjko_16,lloyd_16}. A class of quantum machine learning~\cite{harrow_09,clader_13,pan_14,wang_17}, includes quantum data fitting~\cite{wiebe_12,schuld_16,wang_17} and quantum SVM~\cite{rebentrost_14,li_15}, is based on the HHL algorithm~\cite{harrow_09} that can realizes the matrix inversion with exponential speed-up.  To further exploit the power of machine learning on a quantum computer, it is desirable to incorporate nonlinearity~\cite{rebentrost_14,mitarai_18,schuld_18-frAQB,havlicek_18}, without loss of quantum advantages such as significantly speed-up. This subject has been explored from different aspects. One is by quantum neural~\cite{schuld_14,cao_17} that can realize nonlinear transformation of data. However, it is intrinsically difficult due to the unitary(thus linear) nature of quantum operations.  Another recently developed approach~\cite{mitarai_18,schuld_18-frAQB,havlicek_18}, using variational quantum circuits, instead introduces nonlinearity by encoding classical data in the parameters of quantum gates. Yet quantum advantages from this approach are awaited to be proven, and the model is also lack of interpretability. An approach we adopt here refers to linear models, for which the desired nonlinearity is introduced by proper feature maps in the encoding. This method has been applied in quantum support vector machine for classification~\cite{rebentrost_14, chatterjee_17}. 

In this paper, we study quantum algorithms for nonlinear regression, which is another important machine learning task. The implementation is based on a hybrid quantum computer, exploiting the best of both discrete and continuous variables~\cite{furusawa2011quantum,andersen_15,lloyd00,van_08,proctor_17}. The hybrid way is not only desirable for feature maps, for its great capacity to encode huge information, thanks to the exponential large Hilbert space of $n$-qubit system and the infinite dimension~\cite{liu_16,lau_17,schuld_18} of continuous variables, but also is efficient and feasible~\cite{lau_17, zhang_18} for implementing quantum algorithms of machine learning. Even with explicit feature maps, their quantum implementation for regression can be efficient in runtime. As concrete examples, we propose feature maps that can realize kernel ridge regressions, for both polynomial and Gaussian kernels. Our results show that the runtime is $O(\log{MN})$ for the case of polynomial kernel, the same as linear regression, providing exponential speed-up both in the dimension $N$ of data and the number $M$ of samples. For the case of Gaussian kernel, the runtime scales as $O(\log{M})$, based on that quantum random access memory~\cite{giovannetti_08}(qRAM) can efficiently access hybrid quantum states consisting of both qubits and qumodes. As an outlook, we explore novel feature maps using quantum evolutions that may be helpful for predicting physical properties of a system.

%Quantum machine learning may better adopt the wide model approach for the near term application. The main reason is that complicated nonlinear function is hard to be realized in the quantum circuit. It is better to introduce nonlinearity at the encoding step, and the machine learning can be a simple one, as quantum versions of some simple models have been proposed which demonstrate exponentially speed-up.

The paper is organized as follows. We first discuss in Sec.\ref{encoding} encoding schemes, which specify the way of encoding classical data into quantum states, and show how feature maps can be implemented by encoding. After introducing linear regressions as well as their quantum versions in Sec.\ref{sec:kernel regression}, we proposed quantum algorithms for kernel ridge regression with polynomial kernel and the more widely-used Gaussian kernel, respectively. Lastly, we suggest quantum evolution as a feature map and discuss its application.

%In this sense, rather than considering input step of quantum computing  a bottleneck, we take it as an opportunity for expanding the capacity of quantum machine learning from the aspect of encoding.
\section{Encoding schemes and feature maps}
\label{encoding}
% I should discuss the disadvantage and opportunities of encoding scheme. 
In order to process classical/quantum information in quantum setup, the first step is to encode the classical/quantum  information in quantum systems, typically as quantum states. In classical quantum computers, all information is encoded as strings of bits. In quantum systems, information can be encoded in various ways~\cite{schuld_18}, not only in qubits but also continuous variables, e.g., qumodes. Moreover, quantum superposition allows to encode $2^n$ numbers in only $n$ qubits. The choice of encoding scheme greatly affects the following designing of quantum procedure to process information (for example, see a discussion on quantum imagine processing~\cite{yao_17}), and also is the key to introduce nonlinearity. Thus it is desirable for a careful choice of encoding scheme for a specified task. For our purpose we consider how to encode an N-dimensional vector $\vec{a}=(a_1,a_2...a_N)$, $a_i\in R^N$. For illustration we also mention the encoding for a figure with $m_1\times m_2=N$ pixels with each being either white or black, e.g., $a_i=0,1$.  In our notation, classical data $\vec{a}$ is encoded as a quantum state $\ket{\psi_{\vec{a}}}$.  We consider the following two kinds of encoding:

\begin{enumerate}
	\item \emph{Basic encoding} Each component of a vector is encoded into a mode, leading to a product quantum state. It naturally represents a figure as $\ket{\psi_{\vec{a}}}=\otimes_i\ket{a_i}$ using $N$ qubits, and in formula it is the same as a string of bits. For a vector $\vec{a}$ with real number $a_i$, it is unrealistic to encode $a_i$ by a string of qubits, while continuous variables are proper. For instance, one can encode $a_i$ as a coherent state $\ket{a_i}_c$. The quantum state encoding $\vec{a}$ is then an N-mode product coherent state $\ket{\psi_{\vec{a}}}_c=\otimes_i\ket{a_i}_c$. It is also possible to encode $a_i$ directly into a qumode, either position or momentum. Taking the case of position as an example, then $\ket{\psi_{\vec{a}}}_q=\otimes_i\ket{a_i}_q$ is a wave packet locating at position $\vec{a}$ in space. 	
	\item \emph{Amplitude encoding}. The components of a vector are encoded as amplitudes for corresponding quantum basis, learning normally to entangled states. An $n$-qubit quantum system can encode a $2^n$ dimensional vector, demonstrating the power of superposition in quantum world. Assuming normalized $\vec{a}$, then $\ket{\psi_{\vec{a}}}=\sum_{i=1}^{N}a_i\ket{i}$, where $i=i_1i_2..i_{[\log N]}$ is the binary representation. In our notation we directly write it as $\ket{\vec{a}}$. The same procedure is also applied for encoding a figure, leading to a quantum state whose amplitudes take either $0$ or $1/\sqrt{MN}$. It is noted that amplitude encoding scheme can also be adopted using qumodes which are continuous variables~\cite{lau_17}. 
	
\end{enumerate}

What is a consequence due to the encoding, as there is a change of representing information?  A simple and important issue concerns with the similarity between a pair of vectors, namely $(\vec{a}^{(m_1)})^T\vec{a}^{(m_2)}$, and that between two corresponding quantum states after encoding, in the form of inner product of two quantum states
$\braket{\psi_{\vec{a}^{(m_1)}}}{\psi_{\vec{a}^{(m_2)}}}$. The later can be taken as a kernel.  For basic encoding with qubits, whenever overlapping on one qubit is zero, e.g., $\braket{\psi_{\vec{a_i}^{(m_1)}}}{\psi_{\vec{a_i}^{(m_2)}}}=0$, the inner product is zero, even when $(\vec{a}^{(m_1)})^T\vec{a}^{(m_2)}\neq 0$. As a consequence, two images encoded in such a way are orthogonal even when they only differ in a single pixel.
This may be improper for some applications. While encoded as a product of coherent state, the inner product turns out to be $e^{-||\vec{a}^{(m_1)}-\vec{a}^{(m_2)}||^2/2}$, which is a widely-used Gaussian kernel. Thus, an encoding of $\vec{a}$ as a product of coherent state can be utilized as an useful feature map. Remarkably, this feature map to infinite dimensional space is feasible in a physical way due to the infinite dimensionality of quantum states of continuous variables.

For amplitude encoding, it can be verified that the similarity is exactly preserved.
Moreover, polynomial kernels can be easily designed by preparing ~\cite{rebentrost_14} $\ket{\psi^d_\vec{a}}=\ket{\vec{a}}^{\otimes d}$, which can be verified from the inner product  $\braket{\psi^d_{\vec{a}^{(m_1)}}}{\psi^d_{\vec{a}^{(m_2)}}}=(\braket{\vec{a}^{(m_1)}}{\vec{a}^{(m_2)}})^d$.  Such a state introduces d-order features, that is,  $a_{i_1}a_{i_2}...a_{i_d}$. Taking $d=2$ as an example, it maps $\vec{a}$ to $N^2$-dimensional vector $(a_1^2,a_2^2...a_N^2, a_1a_2, a_1a_3,...,a_{N-1}a_N)$, including cross terms $a_ia_j$. A more widely used polynomial kernel is $k(\vec{a}^{(m_1)},\vec{a}^{(m_2)})=((\vec{a}^{(m_1)})^T\vec{a}^{(m_2)}+c)^d$. It equals to map $\vec{a}$ to $(N+1)^2$-dimension vector  $(a_1^2,a_2^2...a_N^2, a_1a_2, a_1a_3,...,a_{N-1}a_N, ca_1,ca_2,...,ca_N,c^2)$. The inclusion of all zero, first and second orders of $a_i$ can provide a more complete description of features that may be required for machine learning. This can be achieved by a modification of the amplitude encoding scheme,$
\ket{\psi_{\vec{a}}}=(c\ket{0}+\sum_{i=1}^{N}a_i\ket{i})/(c^2+||\vec{a}||^2).
$

\section{Nonlinear regression and its quantum version}
\label{sec:kernel regression}
In this section, we first introduce nonlinear regression that uses feature maps, and reveal its relation to kernel ridge regression. Then, we propose a general framework for quantum nonlinear regression. 

\subsection{Nonlinear regression using feature maps}
Given a training dataset of $M$ points $\{\mathbf{a}^{(m)},y^{(m)}\}$, where $\mathbf{a}^{(m)} \in R^N$ is the vector of $N$ features and $y^{(m)}\in R$ is the target value, the goal is to predict $\tilde{y}$ for new data $\tilde{\mathbf{a}}$. We first apply a feature map for each $\mathbf{a}^{(m)}$ to $\phi(\mathbf{a}^{(m)})$, which may be a vector denoted as $\vec{z}^{(m)}$, or a function of continuous variables $x$ denoted as $\phi(\mathbf{a}^{(m)},x)$. We then consider a linear model with parameters $w$ that predicts as an inner product  $\tilde{y}=\innerproduct{w}{\phi(\mathbf{\tilde{a}})}$. When $w$ is a function, it is a functional linear regression~\cite{muller_05,morris_15}. We remark that the regression is linear with $\phi(\mathbf{a})$ but can be nonlinear with the original data $\vec{a}$, and thus is nonlinear regression.  By minimizing the loss function of least-squares errors with $L_2$ regularization,
\begin{equation}\label{eq:min_loss}
w = \text{min}_w \sum_{m=1}^{M}(\innerproduct{w}{\phi(\mathbf{a}^{(m)})}-y^{(m)})^2+\chi||w||^2.
\end{equation} 
Here $||w||^2=\innerproduct{w}{w}$. The regularization term $\chi||w||^2$ makes a constraint on the parameters, and must be required for the functional linear regression, due to the infinite dimension of features.  Solving Eq.~(\ref{eq:min_loss}), $w$ is obtained analytically as $w=\bar{\vec{A}}^T\mathbf{y}$.  Here  $\bar{\vec{A}}=(K+\chi I)^{-1}\vec{A}$,  $\vec{A}=[\phi^T(\mathbf{a}^{(1)}),\phi^T(\mathbf{a}^{(2)}),...,\phi^T(\mathbf{a}^{(M)})]^T$, $I$ represents the identity matrix, and $K$ is a $M\times M$ covariance matrix with elements
$K_{m_1,m_2}=\innerproduct{\phi(\mathbf{a}^{(m_1)})}{\phi(\mathbf{a}^{(m_2)})}$.

The prediction can be rewritten as 
\begin{equation}\label{eq:prediction}
\tilde{y} = \vec{y}^T(K+\chi I)^{-1}\innerproduct{\vec{A}}{\phi(\mathbf{\tilde{a}})}.
\end{equation} 
Note that $\innerproduct{\vec{A}}{\phi(\mathbf{\tilde{a}})}$ is a $M$-dimension vector with $m$-th element $\left\langle \phi(\mathbf{a}^{(m)}),\phi(\mathbf{\tilde{a}})\right\rangle$.  
The wisdom of kernel method is that one can directly calculate the so-called kernel function  $k(\vec{a},\vec{b})=\innerproduct{\phi(\vec{a})}{\phi(\vec{b})}$ instead of making explicit feature maps.  

The covariance matrix $K$ can be calculated in advance. Then in each prediction it requires $M$ times to get all $\innerproduct{\phi(\mathbf{a}^{(m)})}{\phi(\mathbf{\tilde{a}})}$. As it is known that a quantum algorithm for linear algorithm  scales with $\log M$, one may wonder if such a speed-up holds when considering feature maps, which is useful for big data problems with large $M$. 
 
\subsection{Quantum version of nonlinear regression}
We first formulate how a regression can be converted into a quantum task. Then,
quantum algorithm to solve this task is given. Our procedure decomposes the quantum nonlinear regression into two parts: an encoding scheme to load classical data into a quantum state that at the same time realizes a feature map, e.g., $\vec{a} \rightarrow \phi(\vec{a})$; a quantum algorithm that implement linear regression, corresponds to Eq.~(\ref{eq:prediction}). This provides a general framework, and details of implementation would be discussed for specified feature maps. 

\subsubsection{Convert to a quantum task}
Solving classical problems with quantum algorithms exploits the capacity of encoding huge information in quantum systems and the intrinsic quantum parallel way of processing information. It demands a proper way of mapping a classical problem into a quantum task.
To do this for regression, let us firstly analyze the mathematical structure of Eq.~(\ref{eq:prediction}) using single value decomposition(SVD)~\cite{schuld_16}.  The key point is: if we write $\vec{A} = \sum_i \lambda_i \mathbf{u}_i\phi_i^T$ through SVD, where $\mathbf{u}_i^T\mathbf{u}_j=\delta_{ij}$ and $\innerproduct{\phi_i}{\phi_j}=\delta_{ij}$, then it follows $\bar{\vec{A}}=(K+\chi)^{-1}\vec{A} = \sum_i \frac{\lambda_i}{\lambda_i^2+\chi} \mathbf{u}_i\phi_i^T$. Here $\lambda_i$ are singular values of $A$ with corresponding left(right) eigenstate $\mathbf{v}_i$($\phi_i$). The prediction turns to be
$\tilde{y} = \sum_i \frac{\lambda_i}{\lambda_i^2+\chi} \mathbf{u}_i^T\mathbf{y}\innerproduct{\varphi_i}{\phi(\tilde{{\mathbf{a}}})}$.

We assume that classical data is encoded into quantum state
at the encoding step as $\vec{a}\rightarrow \ket{\psi_\vec{a}}$. The details will be presented for specified encoding schemes. This realizes a feature map, where $\braket{x}{\psi_\vec{a}}=\phi(\vec{a},x)$ for continuous variables and $\braket{i}{\psi_\vec{a}}=\vec{z}_i$ for discrete variables. To get a quantum version of Eq.~(\ref{eq:prediction}), we can encode $\vec{A}$ as a quantum state $\ket{\psi_{\vec{A}}}=\sum_m ||\phi(\vec{a}^{(m)})||\ket{m}\ket{\psi_{\vec{a}^{(m)}}}$. Note from Schmidt decomposition we have $\ket{\psi_{\vec{A}}}=\sum_i \lambda_i\ket{\mathbf{u}_i}\otimes\ket{\phi_i}$. The quantum task thus is to get the target state $\ket{\psi_{\bar{\vec{A}}}}=\sum_i \frac{\lambda_i}{\lambda_i^2+\chi}\ket{\mathbf{u}_i}\otimes\ket{\phi_i}$, from the initial state $\ket{\psi_{\vec{A}}}$. 
It can be seen that the transformation maps those coefficients from ${\lambda_i}$ to $\frac{\lambda_i}{\lambda_i^2+\chi}$, which is denoted as singular value transformation. After $\ket{\psi_{\vec{\bar{A}}}}$ is obtained, the prediction is just an inner product between $\ket{\psi_{\vec{\bar{A}}}}$ and the reference state $\ket{\psi_R}=\ket{\mathbf{y}}\otimes\ket{\psi_{\tilde{\mathbf{a}}}}$, where $\ket{\mathbf{y}}=\sum_i y_i\ket{i}$ using the amplitude encoding scheme. 

% While $\ket{\psi_{\vec{A}}}$ and $\ket{\psi_{\mathbf{y}}}\otimes\ket{\phi({\tilde{\mathbf{a}}})}$ can be prepared with normalized $\vec{A}$, $\mathbf{y}$ and $\tilde{\mathbf{a}}$ in advance, it can not do such to $\vec{\bar{A}}$ as it is unknown.

\subsubsection{Quantum algorithm}
\label{subsec:quan_algo_LR}
As discussed in the above, the aim of quantum algorithm is to transform the prepared state $\ket{\psi_{\vec{A}}}$ to the target state $\ket{\psi_{\vec{\bar{A}}}}$. 
Although a direct unitary transformation is possible,  we refer to a more efficient way using ancillary modes. Those ancillary modes firstly register singular values using quantum phase estimation, and then perform singular value transformation based on the registered singular values. We adopt a hyrbid quantum computing that uses qubits for encoding $\ket{\psi_{\vec{A}}}$ and qumodes as ancillary modes~\cite{zhang_18}. The hybrid approach allows a more efficient quantum phase estimation. Moreover, regularization can be simply implemented by a controlled-phase gate on qumodes, instead of performing arithmetic on qubit system that requires many qubits to encode float numbers. Remarkably, singular values are encoded in an entangled two-qumode state~\cite{lau_17}, and homodyne detection on two qumode gives rise to the required singular value transformation.
The hybrid approach with qumodes for linear regression has been discussed in detail in Ref.~\cite{zhang_18}, and here we outline the procedures: 
\begin{enumerate}
	\item \emph{State preparation}. Prepare $\ket{\psi_{\vec{A}}}=\sum_m ||\phi(\vec{a}^{(m)})||\ket{m}\ket{\psi_{\vec{a}^{(m)}}}$ using quantum random access memory~\cite{giovannetti_08}. It uses the addressing state $\sum_m\ket{m}$ to access the memory cells storing quantum states $\ket{\psi_{\vec{a}^{(m)}}}$ in training data registers. 
%	$\ket{\psi_{\vec{a}^{(m)}}}$, leading to $\sum_m|\vec{a}^{(m)}|\ket{m}\ket{\psi_{\vec{a}^{(m)}}}\equiv \ket{\psi_\vec{A}}$ .
	\item \emph{Quantum phase estimation}. Construct $U=e^{i\eta K/{\text{Tr}K}\otimes I\otimes\hat{p_1}\hat{p_2}}$ using the density matrix exponentiation method~\cite{lloyd_14,kimmel_17,lau_17}, which requires to access $n$ copies of density matrix $K/{\text{Tr}K}=\text{Tr}_{2}\ket{\psi_\vec{A}}\bra{\psi_\vec{A}}$. Here the partial trace works on the training data registers.
	Perform $U$ on $\ket{\psi_\vec{A}}\otimes\ket{G_{12}}$, where two-mode state $\ket{G_{12}}=\int dp_1dp_2 e^{-(p_1^2+p_2^2)/2s^2}\ket{p_1}_{p_1}\ket{p_2}_{p_2}$ with squeezing factor $s$~\cite{liu_16, lau_17}. 	
	\item \emph{Regularization}. Perform $e^{i\eta \chi \hat{p_1}\hat{p_2}}$ that shifts $q_2$ mode by $\eta\chi p_1$.
	\item \emph{Singular-value transformation}. After homodyne detections of both qumodes with results $q_1=Q_1$ and $q_2=Q_2$, the state now becomes	 
	\begin{equation}
	 \label{eq:appro_target_state}
	 \sum_i \lambda_i B_i(Q_1,Q_2)\ket{\mathbf{u}_i}\otimes\ket{\phi_i}\otimes\ket{Q_1}_{q_1}\ket{Q_2}_{q_2}.
	 \end{equation}
	 where $B_i(Q_1,Q_2) \sim \frac{e^ {-(Q_1^2+Q_2^2)/{2\alpha_i^2s^2}}}{\alpha_i}$
	 at the limit $\alpha_i^2s^4\sim \varepsilon_q^{-1}$. Here $\alpha_i=\eta(\lambda_i^2+\chi)$. Discarding both qumodes since they are disentangled from qubits,  Eq.~(\ref{eq:appro_target_state}) then approximates 
	 $\ket{\psi_{\vec{\bar{A}}}}$ with each coefficient rescaled by $e^ {-(Q_1^2+Q_2^2)/{2\alpha_i^2s^2}}$. 
 	\item \emph{Prediction}. For new data $\tilde{\mathbf{a}}$, the prediction $\tilde{y}$ is an inner product between the target state $\ket{\psi_{\vec{\bar{A}}}}$ and the reference state $\ket{\Psi_R}$，up to a constant factor. The inner product, denoted as $\tilde{y}'$, can be accessed using the method described in Ref.~\cite{rebentrost_14,cai_15}. Construct
 	$\frac{1}{\sqrt{2}}(\ket{0}_a\otimes\ket{\psi_{\vec{\bar{A}}}}+\ket{1}_a\otimes\ket{\psi_R})$ with an ancillary qubit. Then, a projecting measurement of $\ket{+}\bra{+}$ has a success rate $p=\frac{1}{2}(1+\tilde{y}')$. Thus $\tilde{y}'=2p-1$. 
	\end{enumerate} 
	 
It should be emphasized that the finite squeezing is not only practical, but also is helpful for an efficient running of the quantum algorithm. For finite squeezing, homodyne detection of two qumodes can range in an area $ \varepsilon_q$ centered at $Q_1=Q_2=0$, and it is shown~\cite{zhang_18} that the success rate is proportional to $\varepsilon_q^{3/2}$.

	%The inner product can be implemented with swapping test with run time $\log_2(NM)$.
	% The inner product is obtained using swap test scheme. 	

\section{Quantum nonlinear regression}\label{sec:QNR}

We now address quantum algorithms for kernel ridge regressions, including both polynomial kernel and Gaussian kernel. We focus on preparing $\ket{\psi_A}$, which depends on the encoding schemes. Once $\ket{\psi_A}$ is obtained, the quantum algorithm can be implemented following the procedures in Sec.~\ref{subsec:quan_algo_LR} with necessary adjustments.

\subsection{Quantum polynomial kernel regression}
To introduce polynomial kernel, vector $\vec{a}$ is encoded as $\ket{\vec{a}}$ using amplitude encoding with qubits.  Then  prepare $\ket{\psi_\vec{a}^d}=\ket{\vec{a}}^{\otimes d}$.  Without loss of generality we consider $d=2$. and the state reads  $\ket{\psi^2_\vec{a}}=\ket{\vec{a}}\otimes\ket{\vec{a}}$, corresponding to a feature map  $\vec{a} \rightarrow\vec{z}(\vec{a})=\vec{a}\otimes\vec{a}$. 
To apply quantum algorithm in Sec.~\ref{subsec:quan_algo_LR}, we need firstly to load quantum state $\ket{\psi_{\vec{A}}}=\sum_i |\vec{a}^{(m)}|^2\ket{m}\ket{\psi_{\vec{a}}^2}$. This can be implemented using quantum random access memory, with runtime scaling as $\log(MN)$. Write the Schmidt decomposition as $\ket{\psi_{\vec{A}}}=\sum_i \lambda_i \ket{\vec{u}_{i}}\otimes\ket{\vec{v}_{i}}$.   
Since $\vec{A}\vec{A}^T\ket{\vec{u}_{i}}=(\lambda_i)^2\ket{\vec{u}_{i}}$ and $\vec{A}^T\vec{A}\ket{\vec{v}_{i}}=(\lambda_i)^2\ket{\vec{v}_{i}}$, it is possible to register singular values $\lambda_i$ using either $e^{i\eta\vec{A}\vec{A}^T\otimes I\otimes\hat{p_1}\hat{p_2}}$ or $e^{iI\otimes\eta \vec{A}^T\vec{A}\otimes\hat{p_1}\hat{p_2}}$. If $M<N^2$, then the former is favorable for its manipulation on smaller number of qubits; the latter is favorable when $M>N^2$. Once $\ket{\psi_{\bar{\vec{A}}}}$ is obtained using quantum algorithm in Sec.~\ref{subsec:quan_algo_LR}, predictions can be made for new data $\tilde{\vec{a}}$ that should be encoded as $\ket{\tilde{\vec{a}}}\otimes\ket{\tilde{\vec{a}}}$. The runtime scales as $\log{(MN)}$, which is the same to quantum algorithm for linear regression, with the exponentially speed-up for both the dimension of data $N$ and the number of samples $M$.

\subsection{Quantum Gaussian kernel regression}  

We adopt the coherent-state encoding for vector $\vec{a}$. The N-mode product state $\ket{\psi_{\vec{a}}}_c=D(\vec{a})\ket{0}_c=\otimes_i\ket{a_i}_c$ can be prepared with $O(1)$ time, when all $\ket{a_i}_c=e^{a_ia-a_ia^\dagger}\ket{0}_c$ are generated in parallel. In Fock space it corresponds to a feature map $\vec{a} \rightarrow \vec{Z}(\vec{a})=\otimes\vec{z}(a_i)$, where $\vec{z}(x)=\sum_{n=0}^{\infty}e^{-x^2/2}x^n/\sqrt{n!}\vec{e}_n$  is a vector with infinite dimension that takes $\{\vec{e}_n\}$ as the basis. The feature map can  also be written under continuous variables as $\vec{a} \rightarrow \phi(\vec{a},x)=\braket{x}{\psi_{\vec{a}}}_c $. 

% the state preparation does not require a qRAM. 
To work out the quantum algorithm, some issues should be addressed.
The first important issue is to prepare $\ket{\psi_{\vec{A}}}=\sum_m\ket{m}\otimes\ket{\psi_{\vec{a}^{(m)}}}_c$. This can be realized with a conditional quantum operator $U=\sum_m\ket{m}\bra{m}\otimes D(\vec{a}^{(m)})$, performing on $\ket{0}^{\otimes\log_2M}\otimes\ket{0}_c^{\otimes N}$. The runtime is $O(M)$, and is independent of $N$. 
It is noted that $\ket{\psi_{\vec{A}}}$ can be rewritten by Schmidt decomposition, $\ket{\psi_{\vec{A}}}=\sum_i \lambda_i \ket{\psi_{\mathbf{u}_i}}\otimes\ket{\phi_i}$.
Here $\ket{\phi_i}$ is a quantum state of $N$-mode continuous variables, e.g, $\ket{\phi_i}=\int \phi_i(x_1,...,x_N)\otimes_i\ket{x_i}dq_1..dq_N$.  
Secondly, to register singular values, it is better to perform a quantum operation on qubits using $e^{i\eta K\otimes I\otimes\hat{p_1}\hat{p_2}}$, since it works on $\log_{2}M$ qubits instead of $N$-modes of continuous variables.  Lastly, the prediction step requires to prepare a state $\ket{\mathbf{y}}\otimes\ket{\psi_{\tilde{\mathbf{a}}}}_c$.
Runtime is $O(M)$ but is independent of $N$. 

We may speed up the state preparation of $\ket{\psi_A}$ to a runtime of $\log M$ using qRAM. Although the original qRAM is proposed for qubit systems~\cite{giovannetti_08}, it is in principal feasible for the hybrid discrete- and continuous-variable state $\ket{\psi_{\vec{A}}}$. The address state is constructed as $\sum_m\ket{m}$ with qubits. Then, use qRAM to access the memory cell $m$ that stores the continuous-variable state
$\ket{\psi_{\vec{a}^{(m)}}}_c$. This leads to $\ket{\psi_{\vec{A}}}$ with a run time of $O(\log M)$.
Provided that one can effectively prepare $\ket{\psi_{\vec{A}}}$ with qRAM , the runtime scales with $\log M$ and is independent of $N$. 

We wish to pinpoint that classical data $\vec{a}$ can be encoded into a general $N$-mode Gaussian state with other Gaussian gates such as beam splitter, squeezing operator and so on~\cite{weedbrook_12}. This may be useful since interactions between different original features are taken into account.
\section{Discussions and conclusion}
\label{sec: general_encoding}
As discussed in Sec.~\ref{sec:QNR}, generating feature maps with quantum encoding for regression can reproduce well-known kernel ridge regressions. Here we explore its application beyond classical capability. In general, loading classical data into a quantum state can be realized by a quantum circuit parameterized by those classical data. If we take classical data as parameters for a physical system, then the quantum circuit can describe a quantum evolution which outputs a quantum state. While it is hard to directly predict physical properties merely from few parameters describing the system, a quantum state depending on those parameters may reveal more information for a prediction. This inspires us to think that quantum nonlinear regression may naturally fit for a task itself involving quantum mechanics, for instance, to predict the ground (or n-th order excited) state energy 
from atomic configurations of materials. Classical machine learning can directly learn a mapping between atomic configuration and use it for predictions~\cite{behler_07,rupp_12}. 
To leverage the quantum capability, a feature map by encoding the atomic configuration as a quantum state may give a better kernel for predicting the ground state energy. We may consider a feature map as following: 
$\ket{\phi_{\vec{a}}}=e^{iH'(\vec{a})t_0}\ket{0}$. Here the Hamiltonian $H'(\vec{a})$ should be much simpler than the original Hamiltonian for the material but still captures some key features. The quantum evolution $e^{iH'(\vec{a})t_0}$ may be implemented with a quantum circuit.   With this quantum evolution as the encoding scheme, we may apply the regression using quantum algorithm as in Sec.~\ref{subsec:quan_algo_LR}. Runtime depends on the depth of quantum circuit for realizing the feature map, but still scales with $\log M$, provided that qRAM is applicable. Whether it can give a better performance of regression is still an open problem awaiting for further in-depth investigations.

In summary, we have emphasized the importance of nonlinearity in machine learning using quantum algorithms, and propose how to realize nonlinearity in regression task with desired quantum speedup. We have developed novel encoding schemes for nonlinear feature maps. The quantum algorithm is implemented in a hybrid way that exploits both discrete and continuous variables. Two well-known kernel ridge regressions,  polynomial and Gaussian kernels, have been set up with quantum algorithms using novel encoding schemes. Runtime for both scales $O(\log{M})$ with the number of samples in training dataset $M$. Moreover, we have also explored feature maps by quantum evolutions that may be helpful for quantum regression tasks.

\acknowledgments{
We thank Xi-Wei Yao for inspiring discussions on the encoding schemes. This work was supported by the NKRDP of China (Grant
No. 2016YFA0301800), the NSFC (Grants No. 91636218, No.11474153).} 
\appendix

%\bibliography{kernels}

\end{document}